\title{Renormalized Wolfram model exhibiting non-relativistic quantum behavior}
\author{
        Jos\'e Manuel Rodr\'iguez Caballero \\
                External affiliate of the Wolfram Physics Project \\ Director of Caballero Software Inc. \\ Ontario, Canada
}
\date{\today}

\documentclass[12pt]{article}

\usepackage[english]{babel}

\usepackage{blindtext}

\usepackage{amsmath}
\usepackage{hyperref}
\usepackage{lipsum}
\usepackage{amsfonts}
\usepackage{enumitem}
\usepackage{epigraph}
\usepackage{graphicx}
\usepackage{amsthm}

\theoremstyle{definition}

\begin{document}
\maketitle

\begin{abstract}
We show a Wolfram model whose renormalization generates a sequence of approximations of a wave function having the Pauli-x matrix as Hamiltonian.
\end{abstract}

\section{Introduction}

According to the principle of computational irreducibility, \cite{wolfram2002new}, generically, computations that are not obviously simple cannot be made shorter. This result prevents developing a theory powerful enough to verify any possible prediction in a complex system before its execution. Nevertheless, by coarse-graining the complex system, it may be possible to get a new system that is simple enough to be predictable. For example, Navot Israeli and  Nigel Goldenfeld \cite{israeli2006coarse} were able to make predictions in computationally irreducible cellular automata after applying coarse-graining. This technique, known as \emph{renormalization}, was widely used by Didier Sornette et \emph{al.} \cite{sornette1995complex, saleur1996renormalization, anifrani1995universal, saichev2006renormalization, sornette1999renormalization, zhou2003renormalization, sornette1989failure, gluzman2002classification} to make predictions in multidisciplinary fields. For an introduction to renormalization, the author recommends Didier Sornette's book \cite{sornette2006critical}.

In the present work, we propose an alternative approach to the quantum mechanics of the Wolfram model based on renormalization. In this approach, we apply coarse-graining to the structures presented by Stephen Wolfram \cite{wolfram2020class} in his physics project, and we obtain a sequence of approximations of a wave function. In principle, this approach differs from previous attempts of mathematical formalization of quantum mechanics in the Wolfram model \cite{gorard2020some, gorard2021zx, gorard2020zx}. Furthermore, suppose the mathematical model that we are presenting describes the ultimate reality of the universe. In that case, it will be incompatible with `t Hooft model of quantum mechanics \cite{t2016cellular}, since we are working with templates. In contrast, `t Hooft considers that the universe evolves from one ontic state to another ontic state. A physical interpretation of the approach to the Wolfram model that we are proposing is likely to be related to the many-worlds school.

\section{Definition of the Wolfram model}

Fix a positive integer $K$. Consider the alphabet $\Sigma = \{a_0, a_1, a_2,... , a_K\}$. Define a Wolfram model where the states are words over $\Sigma$ and the non-deterministic evolution rule is the set of all concatenations of the argument and any symbol from $\Sigma$, i.e.,
$$
\Omega = \left\{
\begin{array}{c}
w \mapsto w \, a_0 \\
w \mapsto w \, a_1 \\
... \\
w \mapsto w \, a_K \\
\end{array} \right.
$$
where $w$ is an arbitrary word generated by the alphabet $\Sigma$.

For example, for $K = 2$ and initial condition $a_0$, we get the following multiway system (shown until level $3$, starting by level $0$). We omitted the $a$'s in the label of the vertices and only wrote their subindex to improve visibility.

\begin{center}
\includegraphics[width=\textwidth]{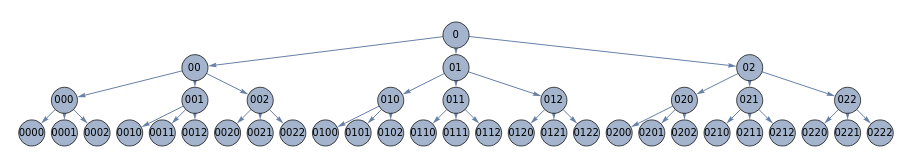}
\end{center}

\section{Renormalization}

Any state $w = w_1 w_2 ... w_{\ell}$, where $w_1, w_2, ..., w_{\ell} \in \Sigma$, of our Wolfram model will be coarse-grained as 

$$
(-i)^m \, \left| m \bmod 2  \right\rangle,
$$
where $m$ is the number of times that the character $K$ appears in the list $w_1, w_2, ..., w_{\ell}$. Notice that the renormalization of $\Omega$ is the multiset of rules

$$
\omega = \left\{
\begin{array}{l}
\left.\begin{array}{c}
|0\rangle \mapsto |0\rangle \\
... \\
|0\rangle \mapsto |0\rangle 
 \end{array} \right\} \text{$K$ times} \\ 
 \left.\begin{array}{c}
|1\rangle \mapsto |1\rangle \\
... \\
|1\rangle \mapsto |1\rangle 
 \end{array} \right\} \text{$K$ times} \\   
   |0\rangle \mapsto -i \, |1\rangle \\
   |1\rangle \mapsto -i \, |0\rangle \\
\end{array} \right.
$$

The multiway system of the renormalized Wolfram model looks as follows (the initial condition and the value of $K$ are the same as before). 

\begin{center}
\includegraphics[width=\textwidth]{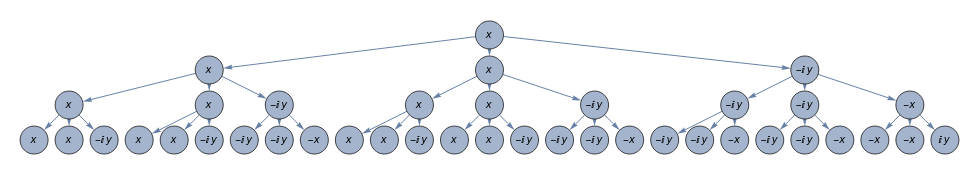}
\end{center}

Instead of $|0 \rangle$ and $|1 \rangle$, we have chosen to write $x$ and $y$ respectively, since the graph is easier to visualize in this way. The initial condition and the value of $K$ are the same as before.

The sum of the elements of the $k$-th level of the multiway system of the renormalized Wolfram model will be called the $k$-th template and will be denoted as $| T_k \rangle$.

\section{Schrödinger equation}
Let $| \Psi(t) \rangle$ be the solution of the Schrödinger equation (for $\hbar = 1$)
$$
i \frac{d}{dt} | \Psi(t) \rangle = H | \Psi(t) \rangle,
$$
satisfying the initial condition 
$$
| \Psi(0) \rangle = | 0 \rangle,
$$
where the Hamiltonian is the Pauli-x matrix $H = \left( \begin{array}{c c} 0 & 1 \\ 1 & 0 \end{array} \right)$. In this case, it is possible to find an explicit expression for the solution,
$$
| \Psi(t) \rangle = \left(\cos t\right) | 0 \rangle - i \left(\sin t\right) | 1 \rangle.
$$

The wave function $| \Psi(t) \rangle$ is the continuous limit of the sequences of normalized templates,
$$
\lim_{K \to \infty} \left(\frac{1}{K^K}\right)^t | T_{\lfloor t K \rfloor} \rangle = | \Psi(t) \rangle.
$$ 

This result is a trivial consequence of the following formula for the exponential of a matrix $M$,
$$
\lim_{n \to \infty} \left(1 + \frac{1}{n} M\right)^n = e^M.
$$

\section*{Conclusions}
Despite the widespread belief that quantum systems cannot be simulated by classical systems, we have shown that, after renormalization, the Wolfram model can be used to approximate a solution of the Schrödinger equation, having the Pauli-x matrix as Hamiltonian. This result motivates the study of the renormalization of Wolfram models as a method to describe the (non-relativistic) quantum systems. Whether there are some advantages in using a discrete template generated by a renormalized Wolfram model instead of a continuous wave function is an empirical question that can be answered by measuring quantum systems and comparing the prediction of the renormalized Wolfram model with those of mainstream quantum mechanics.

\section*{Acknowledgments}
The author would like to thank Stephen Wolfram for the interesting exchange of ideas. Also, the author is very grateful to all the members of the Wolfram Physics Project for having contributed every day to the development of this new approach to fundamental physics.

\nocite{*}

\bibliographystyle{alpha}
\bibliography{mybibfile}

\end{document}